\documentclass{egs}                  

\usepackage{times}                   

\usepackage{amssymb}

\usepackage{graphicx}

\printfigures              

\def\sun{\mbox{$\odot$}}

\begin{document}
\title{The PLANET microlensing follow-up network: Results and prospects
for the detection of extra-solar planets}

%
%
%
%
\author[1]{M. Dominik}
\affil[1]{Kapteyn Astronomical Institute, Postbus 800, 9700 AV Groningen,
The Netherlands}
\author[2]{M. D. Albrow}
\author[3]{J.-P. Beaulieu}
\author[4]{J. A. R. Caldwell}
\author[5]{D. L. DePoy}
\author[5]{B. S. Gaudi}
\author[5]{A. Gould}
\author[6]{J. Greenhill}
\author[6]{K. Hill}
\author[7]{S. Kane}
\author[8]{R. Martin}
\author[4]{J. Menzies}
\author[1]{R. M. Naber}
\author[1]{J.-W. Pel}
\author[5]{R. W. Pogge}
\author[2]{K. R. Pollard}
\author[1]{P. D. Sackett}
\author[9]{K. C. Sahu}
\author[4]{P. Vermaak}
\author[6]{R. Watson}
\author[8]{A. Williams}

\affil[2]{Univ. of Canterbury, Dept. of Physics \& Astronomy, Private Bag 4800,
Christchurch, New Zealand}
\affil[3]{Institut d'Astrophysique de Paris, 98bis Boulevard Arago,
75014 Paris, France}
\affil[4]{South African Astronomical Observatory, P.O. Box 9, Observatory 7935,
South Africa}
\affil[5]{Ohio State University, Department of Astronomy, Columbus, OH 43210, U.S.A.
}
\affil[6]{Univ. of Tasmania, Physics Dept., G.P.O. 252C, Hobart, Tasmania 7001,
Australia}
\affil[7]{School of Physics \& Astronomy, 
University of St Andrews, North Haugh, St Andrews, 
Fife  KY16 9SS, Scotland}
\affil[8]{Perth Observatory, Walnut Road, Bickley, Perth 6076, Australia}
\affil[9]{Space Telescope Science Institute, 3700 San Martin Drive, Baltimore,
MD 21218, U.S.A.}

\date{Manuscript version from 24 October 2001}

\journal{Planetary and Space Science}   
%
%
\firstauthor{Dominik}
\proofs{M. Dominik \\ Kapteyn Astronomical Institute \\
Postbus 800 \\ NL-9700 AV Groningen \\ The Netherlands}
\offsets{M. Dominik \\ Kapteyn Astronomical Institute \\
Postbus 800 \\ NL-9700 AV Groningen \\ The Netherlands}

\msnumber{PSS 154}
\received{}
\accepted{}

\maketitle

%
%
%

\begin{abstract}
Among various techniques to search for extra-solar planets, microlensing has some
unique characteristics. Contrary to all other methods which favour nearby objects, 
microlensing is sensitive to planets around stars at distances of several kpc. These
stars act as gravitational lenses leading to a brightening of observed luminous source
stars. The lens stars that are tested for the presence of planets are not generally
seen themselves. 
The largest sensitivity is obtained for planets at orbital separations 
of 1--10~AU offering the view on
an extremely interesting range with regard to our own solar system and in particular to
the position of Jupiter. The microlensing signal of a jupiter-mass planet lasts typically
a few days. This means that a planet reveals its existence by producing a short signal
at its quasi-instantaneous position, so that planets can
be detected without the need to observe a significant fraction of the orbital period.
Relying on the microlensing alerts issued by several survey
groups that observe 
$\sim\!10^7$ stars
in the Galactic bulge, PLANET (Probing Lensing Anomalies NETwork)
performs precise and frequent measurements on ongoing
microlensing events in order to detect deviations from a light curve produced by a single
point-like object.
These measurements allow constraints to be put on the abundance of planets.
From 42 well-sampled events between 1995 and 1999, we infer that less 
than 1/3 
of M-dwarfs in the Galactic bulge have jupiter-mass companions at 
separations between 1 and 4~AU from their parent star, and that less 
than 45~\% have 3-jupiter-mass companions between 1 and 7~AU. 
\end{abstract}

\section{Introduction}\label{sec:intro}


All of our knowledge about objects outside our own solar system is based
on observations of electromagnetic radiation with different
wavelengths:
infrared, optical, ultraviolet, radio, X-ray, $\gamma$-ray.
In any case, planets are not strong emitters and orbit
stars that are much stronger emitters. For optical wavelengths, most
of the light coming from the planet is reflected light originating
between light from a star and reflected light from a planet is about 
$10^{9}$--$10^{10}$. For infrared wavelengths, this ratio becomes much smaller
due to the
thermal emission of the planet, but it is still $10^{4}$--$10^{6}$.
Therefore, the direct detection of
planets from the emitted or reflected radiation is rather difficult.

For this reason, one has to think about how the presence of the planets yields
observable signals in other objects that are easier to detect. 

Since both planets and their parent stars move around their
common center-of-mass,
one can try to detect the small motion of the luminous parent star rather than observing the
dark planet. Two techniques make use of this effect: The radial
velocity technique observing the one-dimensional radial motion of the star by
determining its velocity via Doppler-shift measurements, and
the astrometric technique observing the two-dimensional transverse motion by
position measurements.

The microlensing technique described here is even more indirect. 
Its basic concept is the observation of a
large number of luminous source stars
to wait for their brightening caused by the gravitational bending of
light by 
intervening compact massive objects that pass close to the line-of-sight and
act as so-called gravitational lenses.
If these {\em lens} objects are surrounded by planets, there is some chance that
their gravitational field causes additional variations in the observed
brightness of the source stars. This means that microlensing can detect
unseen planets orbiting unseen
stars.
Planets orbiting the {\em source} stars and passing in front of them
would not yield to a microlensing signature resulting in a brightening of the
observed stars, but to an occultation resulting in a dimming. Such 
signals can also be used to detect planets.

Among the different techniques, microlensing has some unique 
characteristics.
Since microlensing relies on a chance alignment between source stars and
lens objects, 
there is no opportunity to select the parent stars to be tested for the existence of
planets, contrary to the methods relying on observations
of luminous parent stars.
Contrary to radial-velocity or astrometric searches, there is no need to wait for the orbital period
of the planet. The passage of the star close to the line-of-sight yields a signal
that lasts a few months, while the signal due to the planet is even shorter,
a few days for a jupiter-mass planet.
The favourite range for the detection of planets by microlensing is an orbital
separation of 1--10~AU, i.e. a range comparable to our own solar system, extending
roughly from Earth to Saturn with the most massive planet Jupiter being near its
center, whereas radial velocity searches favour small separations or 
short orbital periods (related to the separations by Kepler's 3rd 
law), and astrometric searches favour large separations or large 
orbital periods, and therefore solar systems that are unlike our own.
Contrary to methods on luminous stars that favour nearby objects, microlensing
favours objects halfway between the observer and the source star, i.e. parent stars at
several kpc distance, and is therefore a unique method to determine the abundance of 
planets around such distant stars.

Here we discuss the prospects of planet detection by microlensing and in particular the
prospects and recent results of our ongoing PLANET experiment.

\section{Microlensing surveys}

As pointed out by \citet{pac86,pac91} and \citet{KP94},
the probability for an alignment of massive compact foreground objects
with luminous source stars in our
Galaxy that yields a significant brightening (30~\%) at a given time
is of the order of $10^{-6}$. One must therefore observe a large number
of stars in order to see a significant number of ongoing 'microlensing events'.
Therefore, fields on the sky with a large number
of stars, such as the Galactic bulge, are of special interest.
About $\sim\!10^7$ stars in the Galactic bulge are currently sampled 
daily (weather permitting) by the 3 collaborations OGLE \citep{OGLE}, EROS 
\citep{EROS}, and MOA \citep{MOA}. From 1995 to 1999, such a survey 
has also been carried out by the
MACHO collaboration \citep{MACHO1,MACHO2}.
These surveys have been equipped with on-line data-reduction systems
so that microlensing events can be caught in real-time.
Issuing alerts on suspected microlensing events by
e-mail and or via designated alert pages on the
web\footnote{OGLE (1994--): {\tt http://www.astrouw.edu.pl/\~{}ftp/ogle/ogle2/ews/ews.html}\\ 
MACHO (1995--1999): {\tt http://darkstar.astro.washington.edu/}\\ 
EROS (1998--): {\tt http://www-dapnia.cea.fr/Spp/Experiences/EROS/alertes.html} \\
MOA (2000--): {\tt http://www.phys.canterbury.ac.nz/\~{}physib/alert/alert.html}} allows
follow-up observations of these events to be undertaken.
The number of publicly issued alerts in the recent years is shown in 
Table~\ref{alerttable}.

\begin{table}
\caption{Number of public alerts issued by the 4 collaborations OGLE, 
MACHO, EROS, and MOA in the different observing seasons from 
1994--2000. Because some alerts correspond to the same event, the number of 
total alerts is sometimes smaller than the 
sum of alerts of the different collaborations. Alerts are regarded 
as {\em useful} for peak magnifications $A_0 \geq 2$ and 
baseline magnitudes $V_0 \leq 20.5$ (MACHO and EROS), 
$I_0 \leq 19$ (OGLE), or $R_0 \leq 19.5$ (MOA), where alerts 
that show obvious anomalies such as caused by binary lenses or 
sources, have been eliminated. As additional criterion,
{\em high-magnification} alerts fulfill the condition $A_0 \geq 10$
and alerts on {\em bright} stars fulfill
$V_0 \leq 17.5$ (MACHO and EROS), 
$I_0 \leq 16$ (OGLE), or $R_0 \leq 16.5$ (MOA). The entry `---' means 
that the survey was not operational, while the entry `0' means that no 
such alerts have been issued.}
\label{alerttable}
\vspace*{0.5em}
\begin{tabular}{lccccccc}
\multicolumn{8}{c}{all alerts} \\ \hline
 & 1994 & 1995 & 1996 & 1997 & 1998 & 1999 & 2000 \\ \hline
OGLE & 2 & 6 & --- & --- & 41 & 46 & 75 \\
MACHO & --- & 41 & 31 & 69 & 50 & 61 & --- \\
EROS & --- & --- & --- & --- & 4 & 8 & 7 \\
MOA & --- & --- & --- & --- & --- & --- & 13 \\ \hline 
total & 2 & 47 & 31 & 69 & 89 & 110 & 94 
\end{tabular}

\vspace*{0.5em}
\begin{tabular}{lccccccc}
\multicolumn{8}{c}{useful alerts} \\ \hline
 & 1994 & 1995 & 1996 & 1997 & 1998 & 1999 & 2000 \\ \hline
OGLE & 1 & 3 & --- & --- & 25 & 22 & 28 \\
MACHO & --- & 28 & 17 & 37 & 30 & 29 & --- \\
EROS & --- & --- & --- & --- & 2 & 2 & 3 \\
MOA & --- & --- & --- & --- & --- & --- & 7 \\ \hline 
total & 1 & 31 & 17 & 37 & 53 & 51 & 37 
\end{tabular}

\vspace*{0.5em}
\begin{tabular}{lccccccc}
\multicolumn{8}{c}{useful high-magnification alerts} \\ \hline
 & 1994 & 1995 & 1996 & 1997 & 1998 & 1999 & 2000 \\ \hline
OGLE & 0 & 0 & --- & --- & 5 & 6 & 5 \\
MACHO & --- & 2 & 0 & 3 & 1 & 1 & --- \\
EROS & --- & --- & --- & --- & 0 & 0 & 0 \\
MOA & --- & --- & --- & --- & --- & --- & 4 \\ \hline
total & 0 & 2 & 0 & 3 & 6 & 6 & 9 
\end{tabular}

\vspace*{0.5em}
\begin{tabular}{lccccccc}
\multicolumn{8}{c}{useful alerts on bright source stars} \\ \hline
 & 1994 & 1995 & 1996 & 1997 & 1998 & 1999 & 2000 \\ \hline
OGLE & 0 & 0 & --- & --- & 3 & 4 & 5 \\
MACHO & --- & 3 & 0 & 1 & 0 & 2 & --- \\
EROS & --- & --- & --- & --- & 1 & 1 & 1 \\
MOA & --- & --- & --- & --- & --- & --- & 2 \\ \hline
total & 0 & 3 & 0 & 1 & 4 & 7 & 8 
\end{tabular}
\end{table}

\section{The PLANET experiment}
The aim of PLANET (Probing Lensing Anomalies NETwork)
is to perform precise and frequent multi-band observations
of ongoing microlensing events in order to study departures from
a light curve that is due to lensing of a point source by a single
point-like lens. The origin of these departures can be due to
blending of the light of the source star by other stars (in particular the
lens) or due to effects by
binary lenses (including planets), binary and extended sources, 
or the parallax effect due to the
motion of the Earth around the Sun.

With these observations, 
PLANET yields valuable contributions
to the fields of Galactic structure and dynamics, binary stars, extra-solar
planets, stellar atmos\-pheres, and variable stars.

PLANET uses a network of four 1m-class optical telescopes in the southern 
hemisphere with the following detectors:
\begin{itemize}
\item South African Astronomical Observatory (SAAO) 1.0m at Sutherland, South 
Africa: camera with beam splitter and optical $2048 \times 2048$ CCD,
infrared $1024 \times 1024$ CCD, both with $0.3''$ per pixel,
field of view $10' \times 10'$ (optical), $5' \times 5'$ (infrared) 
\item Yale 1.0m at Cerro Tololo Inter-American Observatory (CTIO), Chile:
camera with same specifications as at SAAO
\item Canopus Observatory 1.0m near Hobart, Tasmania, Australia:
optical camera with $512 \times 512$ CCD, $0.43''$ per pixel,
field of view $3.7' \times 3.7'$
\item Perth Observatory 0.6m at Bickley, Western Australia:
optical camera with $512 \times 512$ CCD, $0.60''$ per pixel,
field of view $5.1' \times 5.1'$
\end{itemize}

The distribution of these telescopes in longitude allows 
to monitor our targets in the 
Galactic bulge continuously during our observing season from April to 
September each year.
Our main observing band is I, while we take images in V about half as frequently.
PLANET started its observations with a one-month pilot season in 1995 \citep{PLANET1995} and
went fully operational in 1996. We have taken data on 9 bulge events in 1995, on
21 events in 1996, on 31 events in 1997, on 33 events in 1998, on 36 
events in 1999, and on 20 events in 2000.
During the each of the 1998 and 1999 seasons,
we have collected $\sim\!4000$ I-frames and $\sim\!2000$ V-frames.

Its experimental design allowed PLANET to obtain some
important results outside the field of extra-solar planets discussed here.
The measurement of the
relative proper motion between lens and source in a fold caustic-crossing
event towards the Small Magellanic Cloud (SMC) yielded evidence that the lens is
located in the SMC itself and not in the Galactic halo 
\citep{PLANET:SMC1-2,PLANET:SMC1}. 
Futhermore, PLANET has been able to measure limb-darkening coefficients
for three giant 
stars (at several kpc distance) \citep{PLANETM28,PLANET:M41,PLANET:OB23},
and by combining 
PLANET data with those of other microlensing collaborations, for a 
dwarf star in the SMC \citep{joint}.
By detecting motion effects in a binary lens \citep{PLANET:M41}, 
tight constraints for
the lens mass, distance, and rotation period have been derived.
During the regular PLANET observations at the SAAO 1m, the optical 
counterpart of a $\gamma$-ray burst has been discovered in  
R-band and has been followed over the next two days in R-, V-, and 
I-band \citep{PLANET:GRB}.
Finally, PLANET has been able to observe 
variations in the
H$\alpha$ equivalent width during a caustic crossing event with the 
FORS1 spectrograph at the VLT, giving a powerful test 
for stellar atmosphere models of the lensed K giant \citep{PLANET:EB5alph}.

The current status of the PLANET experiment including a list of currently
monitored events and their parameters as well as recent results can be
obtained from the PLANET webpage
{\tt http://www.astro.rug.nl/\~{}planet}.

\section{The theory behind microlensing}
Microlensing uses the effect of the deflection of light due to the
gravitational field of a massive compact
object. 
If $M$ denotes the mass of this object, and $r$ denotes the separation
of the light ray from it, the light ray is deflected by the angle
\citep{Ein15}
\begin{equation}
\hat \alpha = \frac{4GM}{c^2}\,\frac{1}{r}\,,
\end{equation}
where $G$ is the constant of gravitation, and $c$ is the speed of light. 
This deflection yields two possible light trajectories from the
source to the observer resulting
in two images of the same source object on the sky, where the observed 
brightness of these images differs 
from the intrinsic brightness of the source object.
The characteristic physical dimension of microlensing is given by the
{\em angular Einstein radius}
\begin{equation}
\theta_{\rm E} = \sqrt{\frac{4GM}{c^2}\,\frac{D_{\rm LS}}{D_{\rm L}\,D_{\rm S}}}
\,,
\end{equation}
where $D_{\rm L}$ and $D_{\rm S}$ are the distances from the observer to
the lens and the source, respectively, and $D_{\rm LS}$ is the distance between
lens and source. The angular Einstein radius quantifies the quality of alignment between
lens and source on the sky. If $u$ denotes the angular separation between lens and
source in units of $\theta_{\rm E}$, the total magnification of the source, i.e. the 
sum of the magnifications of the two images reads
\begin{equation}
A(u) = \frac{u^2+2}{u\,\sqrt{u^2+4}}\,,
\end{equation}
while the separations between the images is given by
\begin{equation}
\Delta \theta = \sqrt{u^2+4}\,\theta_{\rm E} \geq 2 \theta_{\rm E}\,.
\end{equation}
The smaller the separation between lens and source, the larger the total magnification
$A$; for a separation of
$\theta_{\rm E}$, one obtains $A = 3/\sqrt{5}
\approx 1.34$. For large separations, the magnification falls off as
$A \simeq 1+ 2/u^4$, where the single images contribute
$A_+ \simeq 1+1/u^4$ and $A_- \simeq 1/u^4$, so that 
the total magnification quickly reaches 1 and one image is strongly demagnified.
To yield a significant magnification, the separation therefore cannot
exceed a few $\theta_{\rm E}$, corresponding to a separation
of the images that does not exceed a few $\theta_{\rm E}$ either. 
For large image separations $\Delta \theta \gg \theta_{\rm E}$, the fainter image will escape detection
due to its faintness.

For a source object near the Galactic center and a 
solar-mass lens object halfway between us and the source, one obtains
$\theta_{\rm E} \sim 1~\mbox{mas}$. This means that the images cannot be resolved 
with optical telescopes on the ground (typical resolution $1''$) and also not
with the Hubble Space Telescope (resolution of $0.04''$). Therefore, the only photometric
observable is the total light composed of the two (unresolved) images.

The dynamics of the Galaxy determines the relative proper motion $\mu$ between lens
and source. Therefore, their separation is a function of time
\begin{equation}
u(t) = \sqrt{u_0^2 + \left(\frac{t-t_0}{t_{\rm E}}\right)^2}\,,
\end{equation}
where $t_{\rm E} = \theta_{\rm E}/\mu$ denotes the times in which the lens moves by $\theta_{\rm E}$ 
relative to the source, $u_0$ denotes the smallest separation between lens and source, and $t_0$ 
denotes the point of time at which $u(t) = u_0$. 
This means that one observes characteristic light curves determined 
by the parameters $u_0$, $t_0$, and
$t_{\rm E}$, where $u_0$ determines the peak magnification of the curve,
which is given by $A_0 = A(u_0)$, and $t_{\rm E}$ denotes the
duration of the 'microlensing event'. 

For source stars in the Galactic bulge, lens objects can be stars in the Galactic
bulge itself or in the Galactic disk. 
A typical time scale is
\begin{equation}
t_{\rm E} \sim 40~\mbox{d}\,\cdot\,\sqrt{M/M_\odot}\,,
\end{equation}{
i.e. the timescale is proportional to the square-root of the lens mass, and is about one
month for a solar-mass lens object, one day for a jupiter-mass object, and one hour for an
earth-mass object.

\section{Microlensing and planets}
\begin{figure}
  \figbox*{}{}{\includegraphics[width=\hsize]{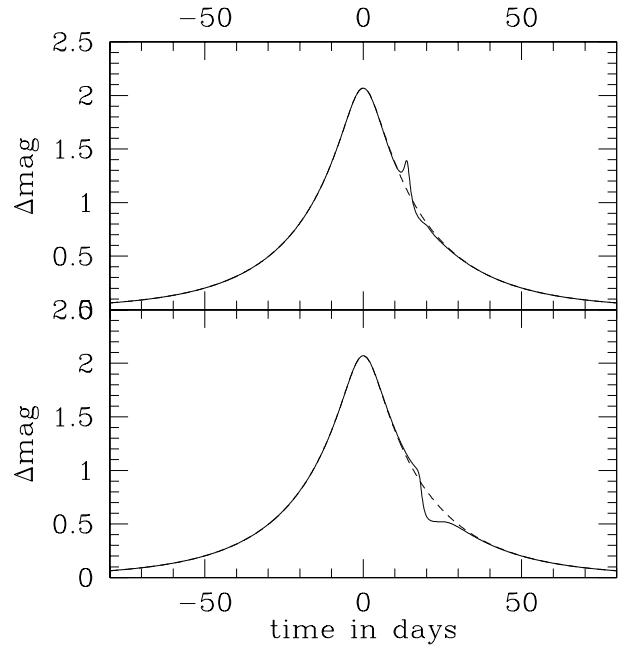}}
  \caption[]{\label{fig:planets}
Planetary perturbations to a light curve caused by microlensing due to a 
star ($t_{\rm E} = 40~\mbox{d}$, $u_0 = 0.15$). The mass ratio is $q=10^{-3}$, resembling
the ratio between Jupiter and the Sun. The upper panel shows a case where
the planet is outside the angular Einstein ring of the star, namely $\theta_{\rm p}
= 1.3\,\theta_{\rm E}$,
while the lower panel shows a case where the planet is inside the angular
Einstein ring of the star, namely at $\theta_{\rm p} = 0.77\,\theta_{\rm E}$.
While in the first case, the
main effect is an increase in amplification, it is a decrease in the second
case.}
\end{figure}

The fact that planets around lens stars cause
observable effects in microlensing light curves was 
first pointed out by \citet{MP91}.
The shape of the light curve is dominated by the effect from the star,
and the effect of the planet can be treated as a perturbation.

However, the planet cannot be modelled as an isolated single lens: the
tidal field of the star at the position of the planet strongly enhances 
the planetary
perturbation by introducing an effective shear \citep{CR79,GL92,GG97,Dom99}.
The strongest effect is obtained if the angular separation $\theta_{\rm p}$ of the planet from
its parent star is comparable to the angular Einstein radius
$\theta_{\rm E}$. To quantify further a range of separations
with strong effects, one
can define a region called the 'lensing zone', defined
by $0.618\,\theta_{\rm E} \leq \theta_{\rm p} \leq 1.618\,\theta_{\rm E}$.\footnote{0.618 means
$(\sqrt{5}-1)/2$ and 1.618 means $(\sqrt{5}+1)/2$.} This also means that the 
probability of seeing a planetary signal is especially large
for planets in this zone compared to outside it. 
The detection probability also depends on the smallest angular 
separation $u_0$ between lens and source and therefore on the peak 
magnification $A_0$.
The probability of detecting a distortion larger than 5\% caused by a jupiter-mass planet in the lensing zone 
around a solar-mass star 
is $\sim\!15\%$  if one considers all events with $A_0 
\geq 1.34$ \citep{GL92}, while it becomes $\sim\!80\%$ 
if only events with $A_0 \geq 10$ are considered
\citep{GriS}.
Examples for perturbations caused by jupiter-mass planets are illustrated in Fig.~\ref{fig:planets}.
For planetary separations $\theta_{\rm p} > \theta_{\rm E}$, the main effect is an increase in magnification,
while for $\theta_{\rm p} < \theta_{\rm E}$, a decrease takes place.

For a lens star at $D_{\rm L} = 4~\mbox{kpc}$, the angular Einstein radius corresponds
to a physical size of
\begin{equation}
r_{\rm E} \sim \sqrt{M/M_\odot}\,\cdot\,4~\mbox{AU}\,,
\end{equation}
so that microlensing has the largest sensitivity for planets in the range 
1--10~AU (taking into account
distributions in the mass of the parent stars and in the distance to the parent stars). If one compares
this distance to our own solar system, one sees that this is quite an interesting range of orbital 
separations: the most massive planet Jupiter lies nearly in the middle of this region, which extends down to
Earth and on the other side nearly reaches Saturn. 

Contrary to the astrometry and the radial velocity methods that rely on signals that are proportional to
the mass of the planet $M_{\rm p}$, 
microlensing can yield signals of the same strength for less-massive planets;
however, the probability of observing this signal decreases, but only 
proportionally to $\sqrt{M_{\rm p}}$.
Nevertheless, a principal
limit is given by the finite size of the source stars.
Only if the angular size of the source star is much smaller than
the angular Einstein radius of the planet, the source star can be approximated
as point-like. Otherwise, its finite source size leads to a reduction of the
planetary signal.
For $M_{\rm p} \sim
10^{-6}\,M_\odot$, the angular Einstein radius of the planet reaches the angular size
of a solar-type source star in the Galactic bulge, so that for jupiter-mass
planets, the point-source approximation is valid. For earth-mass planets,
however, the observed deviation for source stars
with a radius $\sim\!3 R_\odot$
is only $\sim\!1$--2\%, while for larger (giant) stars, there will be no observable signal
at all.

\section{Determining planet parameters}
From a microlensing light curve involving the signal of a planet, only 3
parameters related to the nature of the planet, its parent star, and the
orbit can be extracted: the event time scale $t_{\rm E}$, the mass ratio
between planet and star $q$, and the instantanous projected separation
$d = \theta_{\rm p}/
\theta_{\rm E}$ 
between planet and star in units of Einstein radii \citep{GG97}.
The measured time scale $t_{\rm E}$ is a convolution of
the mass of the star $M$, its distance $D_{\rm L}$,
and the relative lens-source proper motion $\mu$, which are not
known separately.
By assuming a mass density along the 
line-of-sight, a mass spectrum, and a statistical distribution
for the proper motion, one can derive probability distributions for all
quantitities involving
$M$, $D_{\rm L}$,  or $\mu$
for any given observed event with time scale $t_{\rm E}$ \citep{Dom98}.
With reasonable assumptions about the underlying statistical
distributions, the projected separation of the planet from the parent 
star 
can be determined with an uncertainty factor $\sim\!2$, and the mass 
of the planet with an uncertainty factor $\sim\!5$.

Because one is only sensitive to
the instantaneous projection, only a lower limit on the semimajor
axis in units of Einstein radii $\rho = a/r_{\rm E}$ is obtained. Therefore,
the determination of the semimajor axis $a$ depends both on statistics about
the orbits and on statistics about the lens population. The same
is true for the orbital period $P$.
About the eccentricity $\varepsilon$ of the orbit and its inclination $i$, 
microlensing yields no information at all.

\section{PLANET's search for planets}
To be able to characterize the properties of planets, we need to take data
points with a photometric precision of 1--2\% (in order to see 5\% deviations) at
a sampling rate of one point every 1.5--2.5 hours. Though deviations caused
by jupiters last about 1~day, about 10--15 points over such a deviation are
required to be able to extract its true nature. The photometric precision 
determines the exposure time needed for the images and the exposure time
dictates the number of events that can be followed with the needed sampling rate. 
Since the exposure time for a given photometric precision depends on the
brightness of the source stars, the capabilities of PLANET depend
on the brightness of the source stars in the alerted microlensing events.
For giant stars ($V \lesssim 17.5$), exposure times are of the order of
3~min, so that 20 events can be observed at the same time, or 75 events per
season. For fainter stars, the exposure time has to be lengthened to
about 10--15~min, so that around 6 events can be monitored at a given time,
or about 20 events per observing season.
Reliable measurements can be obtained for source stars down to $I \sim 19$, roughly 
corresponding to $V \sim 20.5$ for our targets.

By comparing the capabilities of PLANET with the number of alerted 
events, as shown in Table~\ref{alerttable}, one sees that there
have been up to $\sim 
50$ useful alerts per year, but only 5--10 of them involved giant stars or 
had a peak magnification exceeding 10. Though the total event 
rates from the EROS and 
the MOA surveys are much smaller than those from OGLE and MACHO, there 
is a much larger fraction of alerts with high peak magnification or on 
bright source stars.
While there have been fewer
events on giant stars than can be followed by PLANET, there have been
more events on fainter stars than can be followed. Therefore, among the 
events on fainter stars, we selected those with larger peak 
magnification in order to obtain the strongest possible constraint on 
the abundance of planets, and also observed those events with
anomalous behaviour in order to achieve our science goals outside the 
field of extra-solar planets.

\section{Results}

\begin{figure}
  \figbox*{}{}{\includegraphics[width=\hsize]{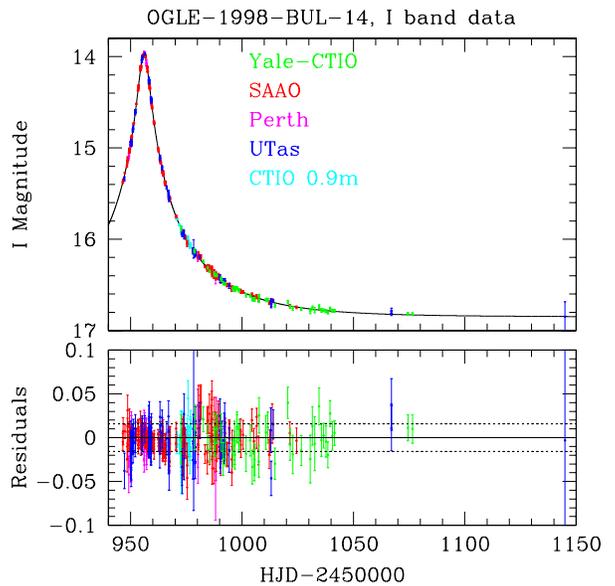}}
  \caption[]{\label{fig:ob9814event}
The PLANET I-band data for the event OGLE 98-Bulge-14. 470 data points have
been taken with 5 telescopes (the 4 telescopes mentioned as PLANET's network
and the CTIO 0.9m). For the peak, the average sampling interval is 
$\sim\!2.5$~h and the typical photometric error bar is $\sim\!1.5$\%.}
\end{figure}

\begin{figure}
  \figbox*{}{}{\includegraphics[width=\hsize]{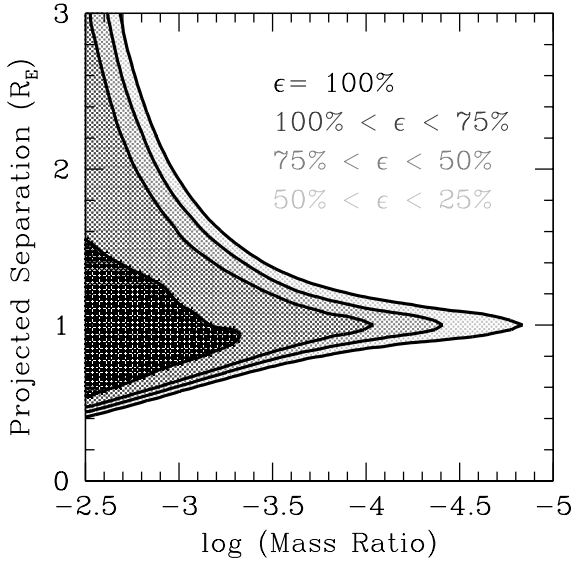}}
  \caption[]{\label{fig:ob9814prob}
Detection efficiency $\varepsilon$ for companions to the parent lens star
that has led to the event OGLE 1998-Bulge-14, for $\Delta \chi^2 = 100$.
The companion is
characterized by its
mass ratio $q$ and its 
instantaneous projected separation $d = \theta_{\rm p}/\theta_{\rm E}$ 
from the parent star in units of Einstein radii. 
The four regions which differ in the shade density correspond to the 
labelled ranges for $\varepsilon$.} 
\end{figure}

Though we would currently expect to detect $\sim\!3$ jupiter-mass planets per year if
every lens star had such a planet within its lensing zone, we have not detected any 
clear planetary signal in our data yet. 
Because of the dense sampling of many microlensing events, constraints on the
presence of planets with certain mass and orbital separation can be derived 
from the fact that no signals have been observed.

For each event, there is a fractional probability that a planet  
would have yielded a detectable signal, namely the 
{\em detection efficiency} $\varepsilon$ which depends on the mass 
ratio $q$ between planetary companion and parent star and the separation 
parameter $d$. \citet{GS99} have developped an algorithm to calculate 
the detection efficiency $\varepsilon(d,q)$ using the criterion for 
a `detectable signal' that
a fit for a binary system (star with planet) yields a $\chi^2$ that exceeds that of 
the best single lens fit by a fixed amount $\Delta \chi^2$.
If the hypothesis of the presence of a companion is rejected 
if no signal is observed, and accepted otherwise, the 
significance level of the corresponding test is 
$\alpha = 1-\varepsilon$.\footnote{However, $\alpha$ does not 
coincide with the probability that a companion is present in the 
system given that no signal has been observed.}

We started our analysis with the well-covered event OGLE 1998-Bulge-14
\citep{PLANET:O14}, 
i.e. the 14th event alerted by the OGLE collaboration
in the 1998 season observed towards the Galactic bulge 
(see Fig.~\ref{fig:ob9814event}). 
This event has a 
timescale $t_{\rm E} \sim 40~\mbox{d}$ and an impact parameter $u_0 = 0.06$, corresponding to
a peak amplification of $\sim\!16$. PLANET has taken 470 data points in I and 139 data
points in V. During the 2 months over the peak, the average sampling interval 
is $\sim\!2.5~\mbox{h}$ and the 
photometric precision is $\sim\!1.5~$\%.

Figure~\ref{fig:ob9814prob} shows the detection efficiency for this 
event where $\Delta \chi^2 = 100$.
The contours shown in the figure reveal the fact that the 
detection efficiency reaches its
maximum for separations around the Einstein radius. 
Assuming $r_{\rm E} \sim 4~\mbox{AU}$,
a planet with Jupiter's mass and a projected separation of Jupiter's orbital separation 
is ruled out.  
The detection efficiency
decreases with smaller mass ratios and drops quickly outside a narrow
region around $\theta_{\rm E}$. Note that this diagram uses the instantaneous projected 
separation in units of Einstein radii. For the conversion to orbital separations, 
three statistical effects have to be considered: the distribution of $r_{\rm E}$, the 
inclination of the orbit, and its phase.

Between 1995 and 1999, PLANET has taken data on 42 well-covered events 
that are suitable for our analysis on the abundance of extra-solar 
planets. The distribution of the 
sampling intervals for this event sample is shown 
in Fig.~\ref{fig:sampling}.

\begin{figure}
  \figbox*{}{}{\includegraphics[width=\hsize]{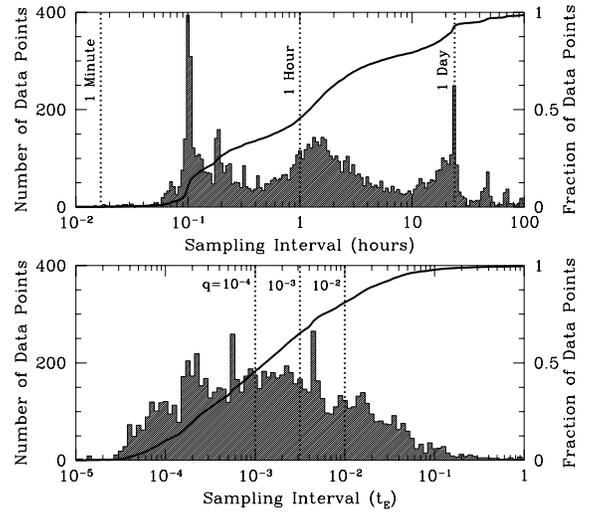}}
  \caption[]{\label{fig:sampling}
Distribution of sampling intervals for the sample of 42 well-covered 
events between 1995 and 1999. The histogram shows the differential distribution
and the solid line shows the cumulative distribution. The upper 
diagram shows the sampling intervals in units of hours, whereas the 
lower diagram shows them in units of the event timescale 
$t_\mathrm{E}$. The vertical dotted lines indicate the approximate 
minimum sampling rates
necessary for detection of companions of the indicated mass ratios.}
\end{figure}

The effective number of lens stars $n_\mathrm{eff}$ that are probed for the existence 
of companions with given characteristics $d$ and $q$ 
is given by the sum of detection efficiencies for all $m$
observed events, 
\begin{equation}
n_\mathrm{eff}(d,q) = \sum_{i=1}^{m} \varepsilon_i(d,q)\,.
\end{equation}
With $f(d,q)$ being the fraction of lens stars with a companion, 
the number of expected signals is
\begin{equation}
N_\mathrm{s}(d,q) = 
f(d,q)\,n_\mathrm{eff}(d,q)\,,
\end{equation}
and the probability for not observing any 
signal in all of the events is given by 
\begin{eqnarray}
P(d,q) & = & 1-\prod_{i=1}^{m} \left(1-f(d,q)\,\varepsilon_i(d,q)\right) \\
& \approx & 1-\exp\left\{-f(d,q)\,n_\mathrm{eff}(d,q)\right\}\,,
\end{eqnarray}
where the approximation holds for $f \varepsilon_{i} \ll 1$.
Therefore, fractions $f(d,q)$ with
\begin{equation}
f(d,q) \geq -\frac{\ln (1-P)}{n_\mathrm{eff}(d,q)}
\label{eq:reject}
\end{equation}
can be rejected at the confidence level $P$. In particular, for $P = 
95~\%$, Eq.~(\ref{eq:reject}) becomes $f(d,q) \geq 3/n_\mathrm{eff}(d,q)$.

\begin{figure}
  \figbox*{}{}{\includegraphics[width=\hsize]{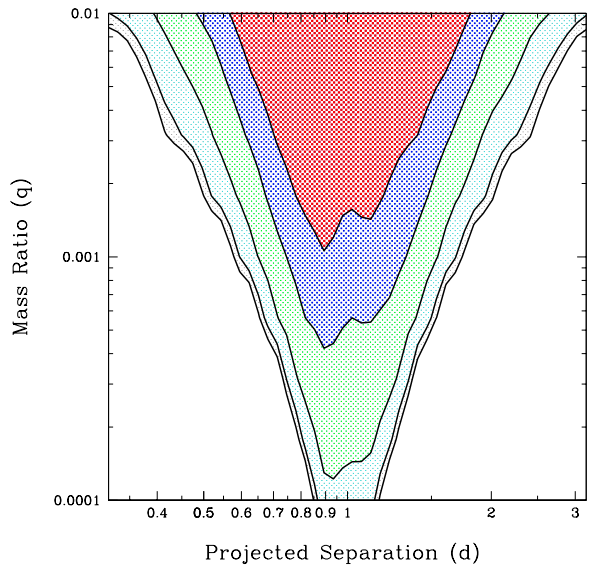}}
  \caption[]{\label{fig:exclude}
Fractions $f(d,q)$ of lens stars having a companion with mass ratio 
$q$ at the separation parameter $d$ which are excluded at 95~\% 
confidence level by PLANET observations on 42 well-covered events 
between 1995 and 1999. From the inside to the outside, the 5 
contours correspond to $f =$~3/4, 2/3, 1/2, 1/3, and 1/4.}
\end{figure}

Our limits on the fractions of systems with planets from our sample of
42 events are shown in Fig.~\ref{fig:exclude}. 
By assuming `typical' values $M \sim 0.3~M_{\sun}$ (corresponding to 
the bulge mass function) and 
$D_\mathrm{L} \sim 6~\mbox{kpc}$, so that $r_\mathrm{E} \sim 2~\mbox{AU}$,
furthermore assuming circular orbits, and averaging over orbital phase 
and inclination, we conclude that less 
than 1/3 
of M-dwarfs in the Galactic bulge have jupiter-mass companions at 
separations between 1 and 4~AU from their parent star, and that less 
than 45~\% have 3-jupiter-mass companions between 1 and 7~AU
\citep{PLANET:fiveshort,PLANET:fivelong}.

\section{The future}
While the termination of the MACHO project at the end of 1999
led to a decrease in the 
number of useful alerts, the 3rd phase of the OGLE 
project starting in 2002 will yield 150--250 useful alerts per 
year, among these 25--40 high-magnification alerts and about the same 
number of alerts on bright source stars. 
This will bring the effective number of lens stars being probed by 
PLANET for jupiter-mass planets in the lensing zone to $\sim\!$~15--25 per year.

An inexpensive pixel-lensing survey towards
the Galactic Bulge which preferably selects events with high 
magnifications \citep{GDP} would yield
$\sim\!50$ alerts per observing season, among them $\sim\!25$ alerts 
with $A_0 \geq 10$ and $\sim\!2$ alerts with $A_0 \geq 200$
\citep{Han:pixelsurvey}, so that PLANET could probe
effectively $\sim\!25$ lens stars per year for jupiter-mass planets and 
$\sim\!3$ lens stars per year for earth-mass planets in the lensing 
zone, the latter essentially coming from the few events with the 
largest peak magnifications. 

This implies that, with three future years of observations, 
PLANET will have 
probed $\sim\!50$--85 stars for jupiter-mass planets, either bringing 
the actual constraint of less than 1/3 of the lens stars being 
surrounded by planets to 3--6\%,
or leading to the detection of up to 8 
jupiter-mass planets per year. Alerts from a possible pixel-lensing 
survey would allow PLANET also to probe $\sim\!3$ lens stars per year for 
earth-mass planets, yielding a constraint that less than 1/3 of the lens stars
have earth-mass planets in their lensing zone if no signals are detected.

Ground-based microlensing searches with advanced technology (and advanced budget) 
could effectively probe $\sim\!10$ stars in the Galactic bulge and disk 
per year for earth-mass planets per year 
and $\sim\!200$ stars per year for jupiter-mass planets
\citep{Pea97,Sac97}. Space-based searches with a dedicated 
satellite (GEST) will even be sensitive to mars-mass planets
and could effectively probe $\sim\!40$ stars per year for
earth-mass planets and $\sim\!2000$ stars per year for jupiter-mass planets 
by microlensing, while detecting a ten times larger number of 
jupiter-mass planets through transits \citep{GEST}.

While the planetary systems studied by ongoing microlensing 
experiments are already at much larger distances than those studied by 
other techniques, microlensing even offers the possibility to constrain the abundance 
of planets around stars in M31, effectively probing $\sim\!35$ lens 
stars for 
jupiter-mass planets in the lensing zone with a network of 
2m-class telescopes \citep{Cov:exgalplanets,BG:exgalplanets,Dom:rdr}.

\begin{acknowledgements}
We thank the survey teams OGLE, MACHO, EROS, and MOA
for providing alerts on ongoing microlensing
events. We also thank the computer staff at Kapteyn Astronomical Institute for their assistence
in developing tools that have improved the communication between the different PLANET sites.
The work of PLANET has been financially supported by
grants AST 97-27520 and AST 95-30619 from the NSF, by grant NAG5-7589 from NASA,
by grant ASTRON 781.76.018 from the Dutch Foundation for Scientific Research (NWO),
by grant Do 629-1/1 from the Deutsche Forschungsgemeinschaft, and by a 
Marie Curie Fellowship (grant ERBFMBICT972457) from the European Union.

\end{acknowledgements}

\balance 

\bibliographystyle{egs}
\bibliography{pss154.bib}       

\end{document}